\documentclass[apjl]{emulateapj}
\usepackage{apjfonts}

\begin{document}
\shortauthors{Fish}
\shorttitle{EVLA Observations of OH Masers in ON~1}
\title{EVLA Observations of the 6035~MH\lowercase{z} OH Masers in ON~1}
\author{Vincent L.~Fish}
\affil{Jansky Fellow, National Radio Astronomy Observatory, P.~O.~Box
  O, 1003 Lopezville Rd., Socorro, NM 87801, vfish@nrao.edu}
\begin{abstract}
This Letter reports on initial Expanded Very Large Array (EVLA)
observations of the 6035~MHz masers in ON~1.  The EVLA data are of
good quality, lending confidence in the new receiver system.  Nineteen
maser features, including six Zeeman pairs, are detected.  The overall
distribution of 6035~MHz OH masers is similar to that of the 1665~MHz
OH masers.  The spatial resolution is sufficient to unambiguously
determine that the magnetic field is strong ($\sim -10$~mG) at the
location of the blueshifted masers in the north, consistent with
Zeeman splitting detected in 13441~MHz OH masers in the same velocity
range.  Left and right circularly polarized ground-state features
dominate in different regions in the north of the source, which may be
due to a combination of magnetic field and velocity gradients.  The
combined distribution of all OH masers toward the south is suggestive
of a shock structure of the sort previously seen in W3(OH).
\end{abstract}

\section{Introduction}

\object[Onsala 1]{Onsala 1} (ON 1) is a massive star-forming region
with an unusual OH maser spectrum.  The ground-state masers, which
have been observed interferometrically several times,
\citep[e.g.,][]{ho83,argon00,fish05,nammahachak06,fish07}, appear in
two disjoint velocity ranges: $< 6$~km\,s$^{-1}$ and
11--17~km\,s$^{-1}$, with no maser emission in between.  This pattern
is reproduced in 6668~MHz methanol emission, which also appears near 0
and 15~km\,s$^{-1}$ \citep{szymczak00}.

The masers presently elude clear interpretation.  Comparison of OH
maser velocities seen in projection against the ultracompact
\ion{H}{2} region \citep[13--14~km\,s$^{-1}$ after correcting for
Zeeman splitting; e.g., from][]{fish05} with the LSR velocity of the
latter derived from a hydrogen recombination line ($5.1 \pm
2.5$~km\,s$^{-1}$) led \citet{zheng85} to interpret the masers as
tracing infall.  However, a proper motion study of the OH masers
suggests that expansion may dominate the kinematics \citep{fish07}.  A
more recent model suggests that the OH masers are associated with a
molecular outflow \citep{kumar04,nammahachak06}.  This model proposes
a shocked molecular torus origin for the southern masers, but
questions remain regarding the overall morphology and velocity
structure of the masers in ON~1.

Few northern star-forming regions have been mapped in the 6035~MHz
line of OH, in part because few radio arrays in the northern
hemisphere have been capable of tuning to the frequency.  A previous
three-station European VLBI Network (EVN) experiment detected seven
6035~MHz maser features in ON~1 but only observed the redshifted
masers \citep{desmurs98}.  Single-dish observations confirm that the
6030 and 6035~MHz masers in ON~1 also appear in two disjoint velocity
ranges, but Zeeman pairing ambiguities have prevented a definitive
measurement of the magnetic field of the blueshifted masers
\citep{baudry97,fish06}.

The upgrade of the National Radio Astronomy Observatory's (NRAO) Very
Large Array (VLA) to the Expanded VLA (EVLA) presents new
observational opportunities in North America
\citep{mckinnon01,ulvestad07}.  Of interest to spectral line observers
is the full frequency coverage between 1 and 50~GHz that will become
available.  This spring, the EVLA for the first time offered
observational capabilities in the extended C-band range of 4.2 to
7.7~GHz, which includes key maser frequencies of OH and methanol.
This Letter reports on initial observations of 6035~MHz OH masers with
the EVLA.

\section{Observations}

The EVLA was used to observe the 6035.092~MHz line of OH toward ON~1
on 2007~May~25 (experiment code AF459).  The array consisted of only
the seven telescopes at the time equipped with the first EVLA C-band
receivers, since the OH line is outside the tuning range of the older
VLA C-band receivers.  The antennas used were distributed primarily
along the western and northern arms of the VLA A-configuration, giving
maximum and minimum baselines of 28.1 and 2.6~km, respectively.  Data
were taken in both left and right circular polarization (LCP, RCP)
using a 781.25~kHz bandwidth divided into 256 spectral channels,
providing a channel width of 0.15~km\,s$^{-1}$ and a velocity range of
39~km\,s$^{-1}$ centered at $v_\mathrm{LSR} = 7$~km\,s$^{-1}$ (Doppler
tracked).  Dual circular correlation products were obtained.

About one hour of usable data were obtained, consisting of 40 minutes
on the masers in ON~1 and 10 minutes on the phase calibration source,
2023+318.  This was also used to set the flux scale (assuming a flux
density of 2.8~Jy from the VLA Calibrator Manual) and for bandpass
calibration.  Blank sky, single-polarization noise in the image plane
with natural weighting was 30~mJy\,beam$^{-1}$ (synthesized beam size
$0\farcs70 \times 0\farcs26$, PA $87\degr$), approximately matching
the expected thermal sensitivity.  Nevertheless, it is possible that
the assumed flux scale is too high by several tens of percent, based
on recent variability seen in the the flux density of 2023+318 at
4.8~GHz from the VLA flux density history database.  The relative
uncertainty in the flux scale between the two circular polarizations
is better than 10\%.

The peak channel of the brightest maser, nearly 50~Jy\,beam$^{-1}$
RCP, was used to determine the absolute position.  The location of
this maser, $20^\mathrm{h}10^\mathrm{m}09\fs090$,
+$31\degr31\arcmin34\farcs91 \pm 0\farcs2$ (J2000), is south and east
of the location obtained from EVN data by \citet{desmurs98}.  The data
in this channel were iteratively self-calibrated (phase-only and
amplitude-and-phase), and calibration solutions were applied to both
polarizations in the entire data set.  Images measuring 30\arcsec\ in
each direction centered on the brightest maser were created in LCP and
RCP.

\section{Results}

\subsection{Maser Locations}

A total of 19 maser features are detected, as listed in
Table~\ref{maser-table}.  Two-dimensional Gaussian components were
fitted to each patch of maser emission above $10\,\sigma$ and grouped
into features in one or more adjacent channels.  One-dimensional
Gaussian components were fitted to the spectrum of each feature to
determine the peak flux density and full width at half maximum (FWHM)
line width.  This was not possible for features detected in fewer than
three channels (indicated by a missing line width in
Table~\ref{maser-table}), so the peak flux and center velocity of the
channel of the brightest maser emission are given for these features
instead.  Taking into account the flux scale caveat, the brightest
6035~MHz maser in ON~1 is approximately as bright or brighter than any
maser seen in the ground-state transitions
\citep{fish05,nammahachak06,fish07}, as noted in W3(OH) as well
\citep{fbs06,fishevn07}.

The locations of the masers are shown in Figure~\ref{map}.  Also shown
are ground-state masers detected with the Multi-Element Radio Linked
Interferometer Network (MERLIN) by \citet{nammahachak06} and with the
Very Long Baseline Array (VLBA) by \citet{fish07}.  The \citet{fish07}
data are aligned relative to the \citet{nammahachak06} data using the
brightest maser detected in the former.  The 6035~MHz masers appear to
have a similar distribution to the ground-state masers near the
continuum region.  The apparent systematic southwardeastward shift of
the 6035~MHz masers relative to the ground-state masers is
significantly less than the beam size and is within the error of
determining the relative alignment between the 6035~MHz and continuum
reference frames.

Position centroids of the masers as measured in consecutive channels
differ at the $0\farcs2$ level but do not shift systematically with
LSR velocity.  This could be due to blending of nearby maser spots at
slightly different velocities or due to position uncertainties caused
by the sparse $uv$-coverage and consequent beam elongation.  The
relative uncertainty in position accuracy between RCP and LCP spots is
negligible ($< 0\farcs01$) in comparison.  Future observations with
the EVLA, when more antennas with extended C-band tuning capability
are available, should provide much better image fidelity, with
correspondingly better determination of maser positions.

The 6035~MHz maser velocities are systemically offset by approximately
0.1~km\,s$^{-1}$ from the \citet{baudry97} Effelsberg observations,
0.2~km\,s$^{-1}$ from the \citet{fish06} Effelsberg observations, and
0.3~km\,s$^{-1}$ from the EVN observations of \citet{desmurs98}.
Accounting for this, masers appear to be brighter on average than in
previous epochs; for instance, the brightest maser is at least three
times as bright as seen by \citet{baudry97} and \citet{fish06}.  The
overall brightening exceeds the errors due to the uncertain flux scale
calibration in these observations and therefore appears to be a real
effect.

\subsection{Magnetic Fields}

\begin{figure}
\resizebox{\hsize}{!}{\includegraphics{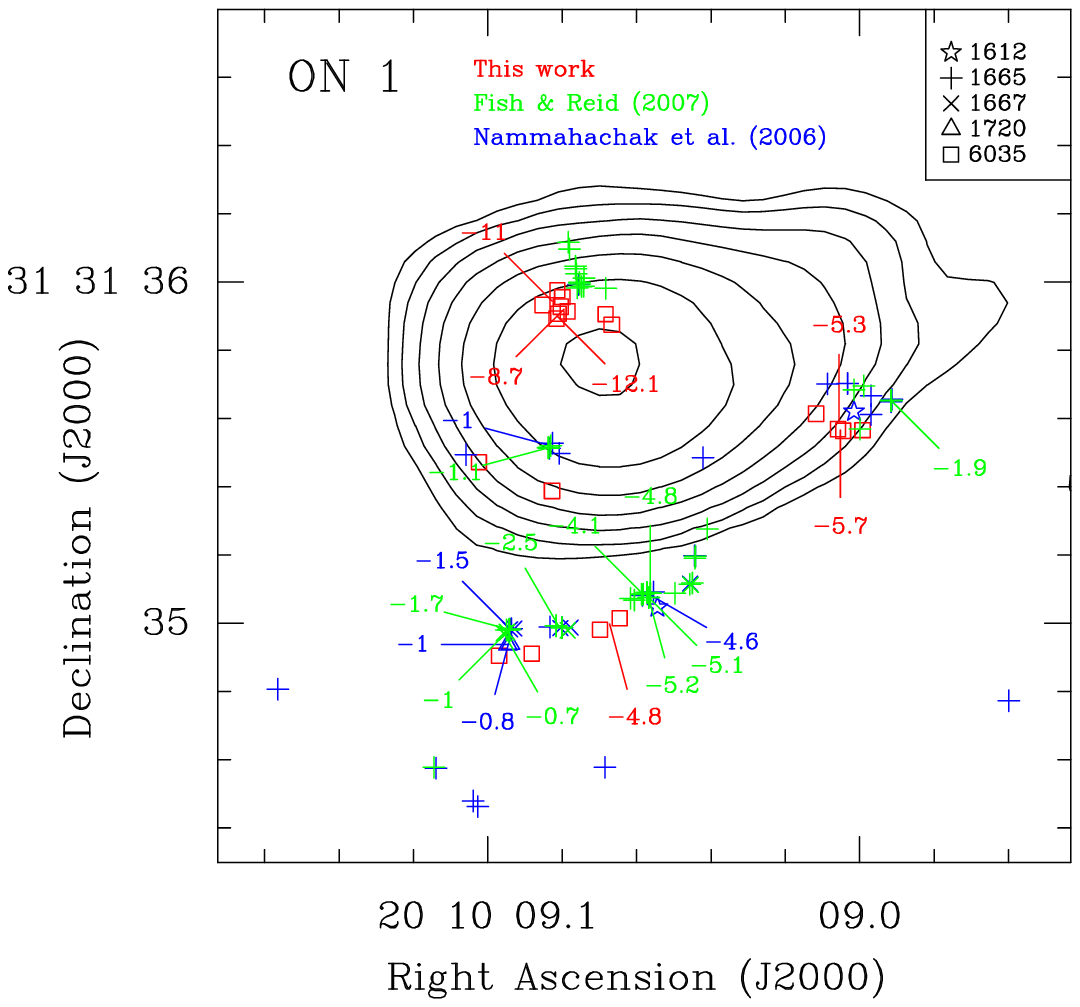}}
\caption{Map of maser emission in ON~1.  Contours indicate 8.4~GHz
  continuum emission from the \citet{argon00} data.  Ground-state
  masers detected by \citet{nammahachak06} and \citet{fish07} are
  shown in blue and green respectively, while 6035~MHz masers (this
  work) are shown in red.  Numbers indicate magnetic fields in
  milligauss obtained from Zeeman splitting at the location pointed to
  by the associated line, color-coded by source.  The small apparent
  southeastward shift of the 6035~MHz masers relative to other maser
  transitions is not significant.
\label{map}
}
\end{figure}

Based on positional coincidence (within the 200~mas uncertainties) of
LCP and RCP maser features at statistically significant different LSR
velocities, six Zeeman pairs are detected at 6035~MHz, listed in
Table~\ref{zeeman-table} and indicated in red on Figure~\ref{map}.
This gives 5 unambiguous pairs as well as a sixth ambiguous pair, as
noted in Table~\ref{zeeman-table}, all consistenly indicating a
negative magnetic field (i.e., one whose line-of-sight projection
points toward the observer).  Three pairs where the velocity
difference is less than the uncertainties are not included; these
could be due to small magnetic fields, linear polarization, or the
pairing of unrelated features.  Assuming that the 6035~MHz masers
should be shifted northward and westward slightly, the $-4.8$~mG
magnetic field agrees with the magnetic field values obtained at
1665~MHz.  This also agrees with a previous Zeeman measurement of this
6035~MHz pair of $-5.3$~mG by \citet{desmurs98}, correcting for the
previously noted velocity offsets.  Magnetic fields of approximately
$-5$~mG are detected on the western side of the continuum emission, in
contrast with a weaker $-1.9$~mG field detected at 1665~MHz.  The
6030~MHz $-4.8$~mG Zeeman pair detected by \citet{fish06} may be
located near the 6035~MHz $-5.7$~mG Zeeman pair at the same corrected
systemic velocity, since \citet{fishevn07} note that 6030~MHz masers
have a strong tendency to appear in very close spatial coincidence
with brighter 6035~MHz masers.  In any case, the magnetic field as
detected by OH masers appears to be negative throughout the source,
although the data do not rule out line-of-sight field reversals on
small scales or where no masers are detected.

In the north, large ($\sim -10$~mG) magnetic fields are detected.  Due
to the large number of maser features in the north, single-dish
measurements have had difficulty pairing the maser features
unambiguously to determine a magnetic field.  \citet{baudry97} claim a
magnetic field of $-2.7$~mG, while \citet{fish06} claim magnetic
fields from $-5.2$ to $-12.8$~mG at blueshifted velocities in the
6035~MHz transition, as well as a $-13.5$~mG Zeeman pair at 6030~MHz,
whose systemic velocity is consistent with the 6035~MHz $-12.1$~mG
Zeeman pair in Table~\ref{zeeman-table}, when correcting for the
0.2~km\,s$^{-1}$ offset.  The larger (absolute) values are also more
consistent with the $-8.3$~mG Zeeman pair detected in the blueshifted
masers in the 13441~MHz highly excited state \citep{fish05b}.  Future
observations at higher angular and spectral resolution may be useful
to confirm and reduce the errors on measurements of the magnetic field
values at 6035~MHz.

The ratio of the fluxes of the bright to weak components of a Zeeman
pair ranges from $\gtrsim 1.0$ to 2.1 at 6035 MHz.  The equivalent
ratio for 1665~MHz masers is 2.1--34.4, while at 1667~MHz two Zeeman
pairs have flux ratios of 2.3 and 2.8 \citep{fish07}.  The two
1720~MHz Zeeman pairs detected by \citet{nammahachak06} each have a
flux ratio of 1.1.  The Zeeman pairs in ON~1 thus fit the general
pattern that the flux ratio between Zeeman components decreases with
decreasing Zeeman splitting coefficient (0.59, 0.354, 0.114\footnote{
\citet{davies74} and \citet{nammahachak06} assume a splitting twice
this at 1720~MHz, appropriate for blending of multiple $\sigma$
components, but a detailed study of the 1720~MHz line profiles in
another source, W3(OH), failed to find evidence of the outer hyperfine
lines \citep{fbs06}.}, and 0.0564~km\,s$^{-1}$\,mG$^{-1}$ at 1665,
1667, 1720, and 6035~MHz, respectively; \citealt{davies74}).

\begin{figure}
\resizebox{\hsize}{!}{\includegraphics{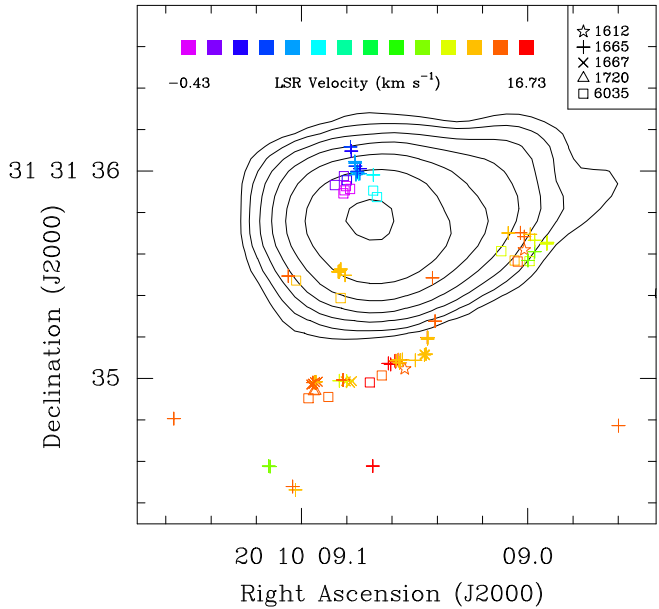}}
\caption{Velocity map of maser spots.  The LSR velocities of maser spots
  from this work, \citet{nammahachak06}, and \citet{fish07} are
  indicated in color.  Northern masers are highly blueshifted compared
  to masers in the south.  Velocities are not corrected for Zeeman
  splitting, which can be significant at 1665 and 1667~MHz.
\label{map-vel}
}
\end{figure}

\begin{deluxetable}{lrrrrll}
\tablecaption{Detected Masers at 6035~MHz\label{maser-table}}
\tablehead{
  \colhead{} &
  \colhead{} &
  \colhead{R.A.} &
  \colhead{Decl.} &
  \colhead{Peak} &
  \colhead{} &
  \colhead{} \\
  \colhead{} &
  \colhead{$v_\mathrm{LSR}$} &
  \colhead{Offset} &
  \colhead{Offset} &
  \colhead{Brightness} &
  \colhead{$\Delta v$} & 
  \colhead{Zeeman} \\
  \colhead{Pol.} &
  \colhead{(km\,s$^{-1}$)} &
  \colhead{(mas)} &
  \colhead{(mas)} &
  \colhead{(Jy\,beam$^{-1}$)} &
  \colhead{(km\,s$^{-1}$)} &
  \colhead{Pair}
}
\startdata
LCP &  $0.06 \pm 0.03$ & $-$105 & 1003 &  1.10 & 0.29    & 1       \\
    &  $0.69 \pm 0.03$ &  $-$78 &  996 &  2.73 & 0.28    & 2       \\
    &  $1.39 \pm 0.08$ &  $-$75 & 1065 &  0.20 & \nodata & \nodata \\
    &  $2.31 \pm 0.03$ &  $-$89 & 1045 &  0.76 & 0.25    & 3       \\
    &  $5.79 \pm 0.08$ & $-$234 &  964 &  0.50 & \nodata & \nodata \\
    & $12.76 \pm 0.08$ & $-$833 &  703 &  0.38 & \nodata & 4       \\
    & $13.96 \pm 0.02$ &    155 &  561 &  2.74 & 0.32    & \nodata \\
    & $14.29 \pm 0.02$ & $-$912 &  653 &  3.67 & 0.36    & 5       \\
    & $14.70 \pm 0.02$ &     96 & $-$6 &  7.83 & 0.24    & \nodata \\
    & $15.65 \pm 0.02$ & $-$200 &   70 &  3.48 & 0.25    & 6       \\
RCP & $-0.43 \pm 0.08$ &  $-$85 & 1017 &  1.98 & \nodata & 1       \\
    &  $0.01 \pm 0.02$ &  $-$72 &  982 &  5.72 & 0.26    & 2       \\
    &  $1.69 \pm 0.08$ &  $-$31 & 1022 &  0.40 & \nodata & 3       \\
    &  $5.72 \pm 0.02$ & $-$216 &  995 &  2.96 & 0.27    & \nodata \\
    & $12.46 \pm 0.08$ & $-$969 &  655 &  0.40 & \nodata & 4       \\
    & $13.97 \pm 0.08$ &  $-$59 &  477 &  6.17 & \nodata & \nodata \\
    & $13.97 \pm 0.08$ & $-$897 &  658 &  6.10 & \nodata & 5       \\
    & $14.69 \pm 0.01$ &      0 &    0 & 49.47 & 0.20    & \nodata \\
    & $15.38 \pm 0.01$ & $-$257 &  104 &  4.57 & 0.27    & 6       
\enddata
\tablecomments{Reference feature location is
  $20^\mathrm{h}10^\mathrm{m}09\fs090$, +$31\degr31\arcmin34\farcs91 \pm
  0\farcs2$ (J2000).  Line widths are FWHM.  Zeeman pairs are listed in
  Table~\ref{zeeman-table}}
\end{deluxetable}

\begin{deluxetable}{lrr}
\tablecaption{Zeeman Pairs at 6035~MHz\label{zeeman-table}}
\tablewidth{\hsize}
\tablehead{
  \colhead{Pair} &
  \colhead{$v_\mathrm{LSR}$\tablenotemark{a}} &
  \colhead{$B$} \\
  \colhead{Number} &
  \colhead{(km\,s$^{-1}$)} &
  \colhead{(mG)}
}
\startdata
1 & $-0.18 \pm 0.08$ &  $-8.7 \pm 1.4 $ \\
2 &  $0.35 \pm 0.04$ & $-12.1 \pm 0.7 $ \\
3 &  $2.00 \pm 0.08$ & $-11.0 \pm 1.4 $ \tablenotemark{b} \\
4 & $12.61 \pm 0.11$ &  $-5.3 \pm 1.9 $ \\
5 & $14.13 \pm 0.08$ &  $-5.7 \pm 1.4 $ \\
6 & $15.52 \pm 0.02$ &  $-4.8 \pm 0.4 $
\enddata
\tablenotetext{a}{Systemic velocity: $0.5*(v_\mathrm{LCP} +
  v_\mathrm{RCP})$.}.
\tablenotetext{b}{While the RCP feature could pair with LCP
  1.39~km\,s$^{-1}$, this would give a magnetic field of +5.3~mG,
  inconsistent in sign and magnitude with other features in the
  region.}
\end{deluxetable}

\section{Discussion}

\subsection{Environment of the Masers in ON~1}

The distributions of the 1665 and 6035~MHz masers are the most alike
of any pair of OH transitions in ON~1.  All the 1665~MHz maser regions
have associated 6035~MHz masers, apart from in the center and in the
extreme south, far from the exciting source where densities are likely
to be low.  The brightest masers at both transitions occur in the line
of maser spots just south of center in Figure~\ref{map}.  This
contrasts with the 1612, 1667, and 1720~MHz masers, which are only
found in the prominent line of masers south of and on the southwestern
limb of the continuum source.  A similar result was seen at VLBI
resolution in W3(OH): 1665 and 6035~MHz masers appear throughout the
source, 1612 and 1720~MHz masers appear only in and near the inner
edge of an apparent shocked torus that is especially well traced by
6.0~GHz masers (including several of the brightest 6.0~GHz masers
detected), and 1667~MHz emission is largely concentrated in areas with
an apparent shock morphology \citep{fishevn07}.  Higher-resolution
observations of the 6.0~GHz masers in ON~1 would be useful to obtain a
better alignment of the 6035~MHz frame with respect to the other
masers and determine whether the southern masers are located
preferentially along the northern edge of the southern shock front.

Some of the 6035~MHz masers may be associated with an outflow, such as
the H$^{13}$CO$^{+}$ oriented northeast-southwest through ON~1
detected by \citet{kumar04}.  \citet{nammahachak06} note that the
southern line of OH masers is oriented nearly perpendicular to this
structure and propose, based additionally on the linear polarization
characteristics of the ground-state masers, that the masers trace a
shock in a confining molecular torus.  Similar conclusions based on
maser distribution and polarization are reached for other OH maser
sources hosting outflows
\citep{hutawarakorn99,hutawarakorn03,hutawarakorn05,hutawarakorn02}.
This interpretation also bears a striking similarity to that proposed
for W3(OH), which hosts very strong 6035~MHz masers \citep{fishevn07},
and may therefore suggest a common mechanism for producing strong
6035~MHz maser emission.

It is probable that there is a coherent, organized structure in the
northern part of ON~1.  Many LCP 1665~MHz masers, including one bright
one, occur in the region, with a few weak RCP masers offset to the
north and west of the LCP emission \citep{fish05,fish07}.  Assuming a
magnetic field of approximately $-10$~mG in the region, LCP and RCP
1665~MHz maser components would be separated by 6~km\,s$^{-1}$ ($20\,
\Delta v$ for an average line width of 0.3~km\,s$^{-1}$) if they both
appear in the region.  (It is likely that weak RCP 1665~MHz masers,
seen near $-2$~km\,s$^{-1}$ in the spectra of \citet{clegg91}, pair
with the dominant LCP masers near +4~km\,s$^{-1}$, but no published
interferometric observations of the ground-state masers have included
the $-2$~km\,s$^{-1}$ features in the observed bandpass.)  If the
magnetic field and velocity gradients are aligned such that the
magnetic field strength decreases (becomes less negative) in the same
direction that the velocity increases, amplification of the LCP
component would be favored \citep[see][]{cook66}.  This effect is much
more pronounced for ground-state OH masers than at 6035~MHz, where the
Zeeman splitting is a factor of 10 smaller in velocity units, such
that the LCP and RCP spectra are separated by $0.6$~km\,s$^{-1}
\approx 2\, \Delta v$ per 10~mG, which is not in excess of the
turbulent velocity component expected in a maser condensation
\citep{reid80}.  Hence, 6035~MHz masers appear in both polarizations
with similar fluxes, while strong LCP 1665~MHz emission appears at
higher velocities (and possibly weaker RCP emission at lower
velocities).

The key question that remains is how the blueshifted masers in the
north connect to the redshifted masers in the south.  It is tempting
to associate both with the HCO$^+$ outflow, but the northern masers
are significantly blueshifted compared to the HCO$^+$ velocity range
of 8--16~km\,s$^{-1}$ \citep{kumar04}.  These authors also note a CO
outflow in the region that spans a larger velocity range, but it is
oriented roughly east-west.  \citet{kumar04} interpret the outflows as
coming from two different sources embedded in the ultracompact
\ion{H}{2} region.  It is possible that the blue and red OH masers are
also associated with two different sources, which would complicate
interpretation of their motions and morphology.  The large ($|B| >
10$~mG) magnetic fields suggest that the northern masers are near a
region of higher density and therefore likely an excitation source.
In W3(OH) the methanol masers in the region of highest magnetic field
strength ($|B| > 10$~mG) have been modelled as undergoing conical
expansion, which \citet{moscadelli02} interpret as possibly being due
to an outflow guided by the helical field from a magnetized disk.

\subsection{Future Directions}

Excluding minor transitional issues, the performance of the EVLA
antennas was as expected.  Early data from the antennas with upgraded
C-band capability are of good quality and are already producing useful
science \citep[e.g.,][]{sjouwerman07}.  As the upgrade continues, more
antennas with 6.0~GHz tuning capability will be added to the array,
providing propotionally better sensitivity and imaging
characteristics.

High spectral resolution VLBI observations of the 1665 and 6035~MHz OH
masers, of the sort obtained for W3(OH) \citep{fbs06,fishevn07}, may
help answer the questions of whether there is an organized velocity
structure in the north and what structure the masers are tracing.  If
the segregated regions of LCP and RCP emission in the north are due to
correlated magnetic and velocity field gradients, small-scale velocity
gradients (i.e., on the size scale of a maser spot) of the northern
masers may show similar magnitudes and position angles.  This would
contrast with W3(OH), in which small-scale velocity gradients show no
correlation except when masers overlap \citep{fbs06,fishevn07}.  Any
such observations of the ground-state masers should have a
sufficiently wide bandwidth coverage to include the $-2$~km\,s$^{-1}$
1665~MHz masers in order to be able to identify magnetic field
strengths (and variations thereof) throughout the northern masers.
While observations of linear polarization exist for the ground-state
masers \citep{nammahachak06}, full-polarization observations at
6035~MHz would help to understand the magnetic field geometry, since
linear polarization vectors are much less subject to being corrupted
by external and internal Faraday rotation at the higher frequency
\citep[e.g.,][]{fishreid06}.

\acknowledgments

The National Radio Astronomy Observatory is a facility of the National
Science Foundation operated under cooperative agreement by Associated
Universities, Inc.  The VLA Calibrator Manual can be found at
\url{http://www.vla.nrao.edu/astro/calib/manual/csource.html}, and the
VLA flux density history database can be accessed at
\url{http://aips2.nrao.edu/vla/calflux.html}.  The author wishes to
acknowledge helpful comments from L.~O.\ Sjouwerman in the manuscript
preparation phase.

{\it Facilities: \facility{EVLA}}


\begin{thebibliography}{}

\bibitem[Argon et al.(2000)]{argon00} Argon, A.~L., Reid, M.~J., \&
  Menten, K.~M.\ 2000, \apjs, 129, 159

\bibitem[Baudry et al.(1997)]{baudry97} Baudry, A., Desmurs, J.~F.,
  Wilson, T.~L., \& Cohen, R.~J.\ 1997, \aap, 325, 255

\bibitem[Clegg \& Cordes(1991)]{clegg91} Clegg, A.~W., \& Cordes,
  J.~M.\ 1991, \apj, 374, 150

\bibitem[Cook(1966)]{cook66} Cook, A.~H.\ 1966, \nat, 211, 503

\bibitem[Davies(1974)]{davies74} Davies, R.~D.\ 1974, Galactic 
Radio Astronomy, 60, 275

\bibitem[Desmurs \& Baudry(1998)]{desmurs98} Desmurs, J.~F., \&
  Baudry, A.\ 1998, \aap, 340, 521

\bibitem[Fish et al.(2006a)]{fbs06} Fish, V.~L., Brisken, W.~F., \&
  Sjouwerman, L.~O.\ 2006a, \apj, 647, 418

\bibitem[Fish \& Reid(2006)]{fishreid06} Fish, V.~L., \& Reid, M.~J.\
  2006, \apjs, 164, 99

\bibitem[Fish \& Reid(2007)]{fish07} Fish, V.~L., \& Reid, M.~J.\
  2007, \apj, in press, arXiv:0708.1186

\bibitem[Fish et al.(2005a)]{fish05} Fish, V.~L., Reid, M.~J., Argon,
  A.~L., \& Zheng, X.-W.\ 2005a, \apjs, 160, 220

\bibitem[Fish et al.(2005b)]{fish05b} Fish, V.~L., Reid, M.~J., \&
  Menten, K.~M.\ 2005b, \apj, 623, 269

\bibitem[Fish et al.(2006b)]{fish06} Fish, V.~L., Reid, M.~J., Menten,
  K.~M., \& Pillai, T.\ 2006b, \aap, 458, 485

\bibitem[Fish \& Sjouwerman(2007)]{fishevn07} Fish, V.~L., \&
  Sjouwerman, L.~O.\ 2007, \apj, in press, arXiv:0706.3352

\bibitem[Ho et al.(1983)]{ho83} Ho, P.~T.~P., Haschick, A.~D., Vogel,
  S.~N., \& Wright, M.~C.~H.\ 1983, \apj, 265, 295

\bibitem[Hutwarakorn \& Cohen(1999)]{hutawarakorn99} Hutawarakorn, B.,
  \& Cohen, R.~J.\ 1999, \mnras, 303, 845

\bibitem[Hutwarakorn \& Cohen(2003)]{hutawarakorn03} Hutawarakorn, B.,
  \& Cohen, R.~J.\ 2003, \mnras, 343, 175

\bibitem[Hutwarakorn \& Cohen(2005)]{hutawarakorn05} Hutawarakorn, B.,
  \& Cohen, R.~J.\ 2005, \mnras, 357, 338

\bibitem[Hutwarakorn et al.(2002)]{hutawarakorn02} Hutawarakorn, B.,
  Cohen, R.~J., \& Brebner, G.~C.\ 2002, \mnras, 330, 349

\bibitem[Kumar et al.(2004)]{kumar04} Kumar, M.~S.~N., Tafalla, M., \&
  Bachiller, R.\ 2004, \aap, 426, 195

\bibitem[McKinnon \& Perley(2001)]{mckinnon01} Mckinnon, M., \&
  Perley, R.\ (Eds.) 2001, "The VLA Expansion Project",
  \url{http://www.aoc.nrao.edu/evla/pbook.shtml}

\bibitem[Moscadelli et al.(2002)]{moscadelli02} Moscadelli, L.,
  Menten, K.~M., Walmsley, C.~M., \& Reid, M.~J.\ 2002, \apj, 564, 813

\bibitem[Nammahachak et al.(2006)]{nammahachak06} Nammahachak, S.,
  Asanok, K., Hutawarakorn Kramer, B., Cohen, R.~J., Muanwong, O., \&
  Gasiprong, N.\ 2006, \mnras, 371, 619

\bibitem[Reid et al.(1980)]{reid80} Reid, M.~J., Haschick, A.~D.,
  Burke, B.~F., Moran, J.~M., Johnston, K.~J., \& Swenson, G.~W.~Jr.\
  1980, \apj, 239, 89

\bibitem[Sjouwerman et al.(2007)]{sjouwerman07} Sjouwerman, L.~O.,
  Fish, V.~L., Claussen, M.~J., Pihlstr\"{o}m, Y.~M., \& Zschaechner,
  L.~K.\ 2007, \apjl{L}, in press, arXiv:0707.3788

\bibitem[Szymczak et al.(2000)]{szymczak00} Szymczak, M., Hrynek, G.,
  \& Hus, A.~J.\ 2000, \aaps, 143, 269

\bibitem[Ulvestad et al.(2007)]{ulvestad07} Ulvestad, J.~S., Perley,
  R.~A., McKinnon, M.~M., Owen, F.~N., Dewdney, P.~E., \& Rodriguez,
  L.~F.\ 2006, \baas, 38, 135

\bibitem[Zheng et al.(1985)]{zheng85} Zheng, X.~W., Ho, P.~T.~P.,
  Reid, M.~J., \& Schneps, M.~H.\ 1985, \apj, 293, 522

\end{thebibliography}
\end{document}